


\def\BulletItem #1 {\item{$\bullet$}{#1}}


\def\AbstractBegins
{
 \singlespace                                        
 \bigskip\leftskip=1.5truecm\rightskip=1.5truecm     
 \centerline{\bf Abstract}
 \smallskip
 \noindent	
 } 
\def\AbstractEnds{\bigskip\leftskip=0truecm\rightskip=0truecm}

\def\ReferencesBegin
{
\singlespace					   
\vskip 0.5truein
\centerline           {\bf References}
\par\nobreak
\medskip
\noindent
\parindent=0pt
\parskip=4pt			        
 }

\def\section #1    {\bigskip\noindent{\bf  #1 }\par\nobreak\smallskip}

\def\subsection #1 {\medskip\noindent{\it [ #1 ]}\par\nobreak\smallskip}



%
 \let\miguu=\footnote
 \def\footnote#1#2{{$\,$\parindent=9pt\baselineskip=13pt%
 \miguu{#1}{#2\vskip -5truept}}}
%
\def\Integers{{\bf Z}}
 

\def\=>{\Rightarrow}
\def\==>{\Longrightarrow}
 
 \def\dal{\displaystyle{{\hbox to 0pt{$\sqcup$\hss}}\sqcap}}
 
%
\def\lto{\mathop
        {\hbox{${\lower3.8pt\hbox{$<$}}\atop{\raise0.2pt\hbox{$\sim$}}$}}}
\def\gto{\mathop
        {\hbox{${\lower3.8pt\hbox{$>$}}\atop{\raise0.2pt\hbox{$\sim$}}$}}}
%
 

\def\isom{\simeq}		




\def\bar{\overline}		



\def\Reals{{\rm I\!\rm R}}	

\def				
  \Complexes
   {{\rm C}\llap{\vrule height6.3pt width1pt depth-.4pt\phantom t}}

\def\tensor{{\otimes}}		



\def\interior #1 {  \buildrel\circ\over  #1}     


\def\semidirect{\mathbin
               {\hbox
               {\hskip 3pt \vrule height 5.2pt depth -.2pt width .55pt 
                \hskip-1.5pt$\times$}}} 

\def\Diff{{\rm Diff}}




\message{...assuming 8.5 x 11 inch paper}

\magnification=\magstep1	
\raggedbottom
\hsize=6.4 true in
 \hoffset=0.27 true in		
\vsize=8.7 true in
\voffset=0.28 true in         
\parskip=9pt
\def\singlespace{\baselineskip=12pt}      
\def\sesquispace{\baselineskip=16pt}      


\def\S{{\cal S}}		
\def\R3{\Reals^3}
\def\RP3{\Reals P^3}



\phantom{}
\vskip -1 true in

\rightline{hep-th/9704081}

\vskip 0.3 true in

\vfill
\bigskip\bigskip

\sesquispace
\centerline{ 
{\bf Geon Statistics and UIR's of the Mapping Class Group}\footnote{*}%
{To appear in Juan Carlos D'Olivo, Martin Klein, Hector Mendez (eds.),
Proceedings of the VII Mexican School of Particles and Fields I Latin 
American Symposium on High Energy Physics Conference, 
held November, 1996, M{\'e}rida, M{\'e}xico (American Institute of Physics).}}
\bigskip

\singlespace			        

\bigskip
\centerline {\it Rafael D. Sorkin}
\medskip
 
\smallskip
\centerline {\sl Instituto de Ciencias Nucleares, 
                 UNAM, A. Postal 70-543,
                 D.F. 04510, Mexico}
\centerline {\sl \qquad\qquad internet address: sorkin@nuclecu.unam.mx}
 

\smallskip
\centerline { \it and }
\smallskip
\centerline {\sl Department of Physics, 
                 Syracuse University, 
                 Syracuse, NY 13244-1130, U.S.A.}
\smallskip
 

\bigskip
\centerline {\it Sumati Surya}
\medskip
\centerline {\sl Department of Physics, 
                 Syracuse University, 
                 Syracuse, NY 13244-1130, U.S.A.}
\smallskip
\centerline {\sl \qquad\qquad internet address: ssurya@suhep.syr.edu}

\AbstractBegins 
Quantum Gravity admits topological excitations of microscopic scale which
can manifest themselves as particles --- topological geons.  Non-trivial
spatial topology also brings into the theory free parameters analogous to
the $\theta$-angle of QCD.  We show that these parameters can be
interpreted in terms of geon properties.  We also find that, for certain
values of the parameters, the geons exhibit new patterns of particle
identity together with new types of statistics.  Geon indistinguishability
in such a case is expressed by a {\it proper subgroup} of the permutation
group and geon statistics by a (possibly projective) representation of the
subgroup.
\AbstractEnds

\sesquispace
\bigskip\medskip

This talk attempts to answer two questions concerning the effect of
topology in generally covariant theories: ``how many free parameters are
there in quantum gravity?'' and ``do topological geons really act like
particles?''  By way of comparison, consider the standard model, which
contains both continuous parameters (like the masses of the quarks, the
strong coupling constant, and $\theta_{\rm QCD}$) and discrete parameters
(like the handedness of the neutrinos, taken relative to $\theta_{\rm
QCD}$, say).  In that flat space theory, all of the many parameters could
be interpreted in terms of the properties of the particles the theory
describes, though in some cases the interpretation would be rather
indirect.  The question here is what additional parameters arise from {\it
spatial topology}, and can they be interpreted in terms of the properties
of the topological particles (geons) that quantum gravity describes?  To
further focus the question for this talk, we will concentrate on the
question of particle {\it{statistics}}.

To our two questions we will encounter the following answers.  There exist
in fact very many additional parameters, both continuous and discrete.  And
the topological excitations show themselves consistently as particle-like,
in the sense that all of the new parameters can be interpreted as telling
us about one of the following:
\BulletItem{internal geon ``qualities'' or ``quantum numbers''}
\BulletItem{geon collision parameters}
\BulletItem{statistics of identical geons}		

\noindent
This positive outcome may be taken to bolster the interpretation of the
topological excitations as particles (geons).

In addition we will see that geons can manifest new types of
indistinguishability and statistics, beyond the (very familiar) bosonic and
fermionic types and the (somewhat less familiar) para-statistical types.
These novel statistics include:
\BulletItem{A statistics based on cyclic subgroups of the permutation
            group $\S_N$}
\BulletItem{Possibly a new statistics based on {\it projective}
          representations of $\S_N$ or its subgroups.}

However, all of these results assume that the topology is ``frozen'', in
the sense that they presuppose a spacetime of the product form
$^3M\times\Reals$.  A key question then is {?`}which quantum sectors will
survive when topology changing processes are incorporated into the theory?.
(Here, we mean by sector a specific set of values of the parameters.  Each
such sector is ``superselected'' in the approximation that topology change
is ignored, but it can be expected to communicate with other sectors in a
more general setting.)

\subsection {Primes and geons}

Before we begin the formal analysis, let us recall the definition of a geon,
and the mathematical decomposition theorem on which it is based.  (For more
about this theorem,  and about the subject of this talk in general, see the
more complete 
exposition in [1], as well as the references therein.)  According
to this theorem, an arbitrary 3-manifold (without boundary) which is
asymptotically $\R3$ can be expressed as
$$
   ^3 M = \R3 \# P_1 \# P_2 \# \cdots \# P_N ,   \eqno(1)
$$
where the $P_i$ are {\it primes} --- manifolds that cannot be built up by
``sticking together'' smaller manifolds.  In fact we will assume further
that the primes are all ``irreducible'', that is, we will exclude the
orientable and nonorientable handles (or ``worm-holes'') from
consideration, as they do not seem to be particle-like in the same sense as
other primes.  By a {\it geon} then, we mean a ``quantized prime'' regarded
as a particle, or in other words, that object in the quantum theory to
which the irreducible prime submanifold corresponds.

To list all possible geons is unfortunately impossible, because there exist
an infinite number of primes, not all of which are known.  Nevertheless, it
is not difficult to visualize an arbitrary geon in a general sort of way:
it is the result of excising a polyhedron from $\R3$, and then performing
appropriate identifications on the boundary faces created by the excision.
One knows that every prime can be made in this manner.


At this point, we must recall also that the presence of non-Euclidean
spatial topology implies the existence of distinct quantum sectors of any
theory that includes gravity.  Moreover, if we assume a fixed spatial
topology, then these sectors do not communicate.  Specifically, if we
assume first that
$$
        ^4 M = \Reals \times M
$$
($M$ being the topology of the spatial slices), and if we assume further
that the meaningful assertions of the theory are all diffeomorphism
invariant (``general covariance''), then we get {\it a distinct quantum
sector for each UIR of the MCG of $M$}, where `UIR' stands for `unitary
irreducible representation' and `MCG' stands for `mapping class
group'.  In fact, the mapping class group of a manifold $M$ (also called
``homeotopy group'' or ``group of large diffeos'') is the analog for
gravity of the group of large gauge transformations in a gauge theory. 
Its formal definition is 
$$
   G =  \pi_0( \Diff^\infty(M) ) := \Diff^\infty(M) / \Diff^\infty_0(M) ,
$$
where $\Diff^\infty(M)$ is the group of diffeomorphisms of $M$ that are
trivial at infinity and $\Diff^\infty_0(M)$ is its connected subgroup.  (An
asymptotically flat approximation should be good for all but cosmological
considerations.)  In order to understand the different quantum sectors, we
thus have to understand the structure of the homeotopy group $G$ and then
to use this information to analyze and interpret the different possible
UIR's of $G$.

\subsection {The structure of the homeotopy group}

In this task a major help is that fact that $G$ is generated\footnote{*}%
{This assertion is fully established in the mathematical literature for the
orientable case, and appears to be true in the nonorientable case as well.
More generally various statements we will make about the structure of $G$
are in some cases known only under the assumption that the Poincar\'e
conjecture is true, or that ``homotopy implies isotopy'' for the primes in
question, or that the primes in question are ``sufficiently large''.  All
our statements are in any case known to be true for large families of
primes, and are plausibly true in general.  If they fail for certain
primes, then the analysis given here will remain true as long as those
primes are absent from the decomposition (1).}
by only three types of diffeomorphism, each with a clear physical meaning.
The three categories of generators are the {\it exchanges}, the {\it
internal diffeomorphisms} and the {\it slides}.  Like every diffeomorphism,
each of these generators can be viewed as the result of a certain {\it
process} (a ``development'' [2]), with the nature of the
process being suggested by the name of the category.  Thus, an exchange is
the result of a process in which two identical primes continuously change
places and a slide is the result of a process in which one prime travels
around a loop threading through one or more other primes, while an internal
diffeomorphism is a diffeomorphism whose support is restricted to a single
prime.  For example, to visualize an exchange of two handles in 2D,
imagine the manifold as a rubber sheet, and imagine taking hold of the
handles and dragging them around until they have changed places.


We spoke just now of ``generators'', but actually the slides and internals are
already complete subgroups of $G$, while the exchanges generate a subgroup
of $G$ isomorphic to the group of all permutations of the identical
(diffeomorphic) primes among themselves.  The basic group theoretical fact
we will need for our analysis is then that
$$
   G = (slides) \semidirect (internals) \semidirect (perms)   \eqno(2)
$$
where the symbol $\semidirect$ denotes semidirect product (with the normal
subgroup on the left).  What this says more concretely is that every
element of $G$ is uniquely a product of three diffeomorphism-classes, one
from each subgroup, and that each subgroup is invariant under conjugation
by elements of the subgroups standing to its right in (2).

\subsection {The UIR's of a semidirect product group}

That fact that $G$ is a semidirect product lets us analyze its UIR's in
terms of representations of its factor groups and their subgroups.  Indeed,
there exists a very general analysis of the UIR's of a semidirect product,
which is explained in detail in [1].  In essence it says the
following.  Let
$$
      G = N \semidirect K
$$
be a semidirect product with $N$ being the normal subgroup.  A finite
dimensional UIR of $G$ is then determined by the following data
 \BulletItem $\Gamma$ \ = a UIR of $N$
 \BulletItem      $T$ \ = a PUIR of $K_0 \subseteq K$ ,

\noindent
where $K_0$ is the subgroup of $K$ that remains ``unbroken by $\Gamma$''.
(Often one calls $K_0$ ``the little group''.)

Perhaps the meaning of ``unbroken'' here can be illustrated most easily
with the classic example of irreducible representations of the Poincar{\'e}
group.  There $G$ is the Poincar{\'e} group itself, $N\isom\Reals^4$ is its
translation subgroup, and the quotient $K\isom{}G/N$ is the Lorentz group.
A choice of UIR $\Gamma$ of $\Reals^4$ is then nothing but a choice of
four-momentum $P^\mu$, and (assuming that $P^\mu$ is timelike) the subgroup
$K_0\subseteq{}K$ left unbroken by this choice is $SO(3)$, the group of
spatial rotations in the center of mass frame of $P^\mu$.  The structure
theorem for UIR's of semidirect products therefore tells us that we get a
UIR of the Poincar{\'e} group (in this case an infinite dimensional UIR) by
choosing a 4-momentum and a UIR $T$ of $K_0=SO(3)$.  In particle language,
we get a UIR by choosing two parameters, a mass\footnote{*}%
{Only the mass counts, according to the theorem, because UIR's $\Gamma$ of
$N$ which belong to the same $K$-orbit (in this case $P^\mu$'s lying on the
same mass shell) yield equivalent representations of $G$.}
and a spin.  In particular the ``internal'' properties of the particle
(it's spin) are determined by the representation $T$ of $K_0$.

For geons the formal situation is analogous, and the key interpretive point
for us will be that the {\it statistics} of the geons will be determined by
a representation of the ``unbroken subgroup'' of the group of permutations
of identical primes.

Finally a comment on the ``P'' which occurs above in the phrase ``PUIR
of $K_0$.''  It stands for ``projective'', and reflects the fact that, even
if one is seeking only ordinary representations of $G$, one may have to
consider projective UIR's of $K_0$, i.e. representations up to a phase.
Whether or not this occurs depends on the case in question; and when it
does occur, the particular equivalence class of projective multipliers
$\sigma$ for $K_0$ which one must use is determined by the properties of
the UIR $\Gamma$ and how $K_0$ acts on it.  (In the case of the
Poincar{\'e} group, it does not occur, which is why we have to consider
just integer spins, unless we want a spinorial representation of the
overall group $G$ itself.)

With these preparations complete, we can proceed to analyze the UIR's of
the MCG of our spatial manifold $M$, and thereby the quantum sectors of
gravity on $\Reals\times{}M$.  One may distinguish two situations,
according to whether the slide subgroup is represented trivially or not.

\subsection {The sectors with trivial slides}

In the simpler case of UIR's $G$ which annihilate the slides, we may give
in effect a complete classification.  In this case, the mathematical
problem is reduced to finding the UIR's of the quotient group,
$G/(slides)$, which by (2) is just the semidirect product
$$
       (internals) \semidirect (perms) .        \eqno(3)
$$
Let us find the finite dimensional UIR's of this group (sometimes called
the ``particle group''), when only a single type of prime $P$ is present in
the decomposition (1).  In this situation the permutation
subgroup is just
$$
   K = (perms) = \S_N ,
$$
the full permutation group on $N$ elements ($N$ being the number of copies
of $P$ which are present), and the group of internal diffeos is the direct
product of $N$ copies of the corresponding group $G^{(1)}$ for a single
prime $P$:
$$
   N = (internals) = G^{(1)} \times G^{(1)} \times \cdots \times G^{(1)} .
$$
From this last equation it follows immediately that the most general UIR of
the normal subgroup $N$ is itself a product, namely the tensor product
$$
    \Gamma = \Gamma_a \tensor \Gamma_a 
              \cdots 
             \Gamma_b\tensor\Gamma_b
              \cdots 
             \Gamma_c .                         \eqno(4)
$$
Notice here that although all the underlying primes are all identical,
there is no reason for all the factors in (4) to be so.  Rather,
we can choose an independent UIR of $G^{(1)}$ for each prime summand, and
in the above formula, the subscripts $a$, $b$, \dots $c$ label the
different equivalence classes among them.  Physically a given $\Gamma_a$
specifies a certain ``internal structure'' for the corresponding geon, and
is therefore a ``species parameter'' or ``quantum number'', analogous in
many ways to the spin of a rigid nucleus or molecule in its body-centered
frame.

With respect to the choice (4),
it is intuitively clear (and true as well) that the unbroken subgroup
$K_0\subseteq(perms)$ reduces to a product of permutation groups,
$$
    K_0 = \S_{N_a} \times \S_{N_b} \times \cdots \times \S_{N_c} ,
    \eqno(5)
$$
where $N_a$ is the number of occurrences of the representation $\Gamma_a$,
$N_b$ of $\Gamma_b$, etc.  The statistics is then given by a UIR $T$ of
$K_0$, that is to say by an independent UIR $T_a$, $T_b,\cdots$ for each
of the subgroups $\S_{N_a}$, $\S_{N_b}, \cdots$.  Each of these $T$'s in
turn, can be specified by a choice of a Young tableau, and determines
whether the corresponding geons will manifest Bose statistics, Fermi
statistics or some particular parastatistics.  Since there is no
restriction on the choice of $T$, there is no restriction on which
combinations of these possible statistics can occur.

Notice that these conclusions (deriving from the structure of $K_0$, as
given in (5)) are entirely consistent with our interpretation
of different choices of internal UIR in (4) as yielding physically
distinct geons --- geons of different ``species''.  In this sense, we can
say that there occurs a {\it quantum breaking of indistinguishability}
conditioned by the choice of representation of $(internals)$.  This
phenomenon, together with the possibility of assigning an arbitrary
statistics to each such resulting species, exhausts the possibilities
inherent in UIR's of the group (3).  Thus, all possible
sectors with trivial slides are accounted for by specifying%
 \BulletItem{ a {\it species} for each geon (i.e. a UIR of $G^{(1)}$) }
 \BulletItem{ a {\it statistics} for each resulting set of identical geons }

\subsection {Some sectors with nontrivial slides}

When the slide subgroup is represented nontrivially, we are unable to give
a full classification of the possible UIR's of $G$, due primarily to the
difficulty of analyzing the UIR's of $(slides)$, but also due in part to
the relative complexity of the manner in which $(internals)$ acts on these
UIR's.  Instead, let us consider a special case which avoids most of these
complications, by choosing a prime that lacks internal diffeomorphisms, and
then limiting ourselves mainly to abelian UIR's of the slides.

The prime in question is $\RP3$, which can be visualized as a region of $M$
produced by excising a solid ball and then identifying antipodal pairs of
points on the resulting $S^2$ boundary.  Since the internal group is
trivial for this prime, we can concentrate on the effects of the slides.
For each pair of $\RP3$'s, one can slide one through the other, with the
square of this slide being trivial (since $\pi_1(\RP3)=\Integers_2$),
making a total of $N(N-1)$ independent order 2 generators.  The complete
group $(slides)$ is then generated by products of these elementary slides,
subject (when $N>2$) to certain geometrically evident commutation
relations, like the fact that slides involving disjoint subsets of the
primes commute with each other.  For abelian representations of $(slides)$,
all of the commutation relations will of course be satisfied trivially.


Since for $\RP3$, $(internals)$ is trivial, the MCG reduces to
$$
          G = (slides) \semidirect (perms)
$$
Hence, according to the general scheme outlined earlier, we get a UIR of
$G$ by choosing first a UIR $\Gamma$ of $N=(slides)$, and then a PUIR $T$
(with the correct projective multiplier $\sigma$) of the resulting unbroken
subgroup $K_0\subseteq{}(perms)$.  As before, we may interpret $K_0$ as
describing the surviving indistinguishability of the geons, and $T$ as
describing the statistics within each set of identical geons.  Here we will
just quote the results of this analysis, referring the listener to
[1] for more details.

\subsection {A single pair of $\RP3$'s} 

\noindent
This case is simple enough that we
can classify all the UIR's of $G$, without limiting ourselves to abelian
representations $\Gamma$. The quantum sectors comprise a 1-parameter
continuous family together with a handful of discrete cases.  Aside from
the long-known violation of the spin-statistics correlation which occurs in
one of the sectors, the intriguing new result is that there exist other
sectors where the Bose-Fermi distinction becomes ambiguous in a certain
sense.  In these sectors the permutation group remains unbroken
($K_0=\Integers_2$), but there is no natural way to say which of its two
representations describes a pair of bosons and which a pair of fermions!
This happens because the UIR's $\Gamma$ of $(slides)$ and $T$ of $(perms)$
mix in such a way that it apparently becomes meaningless to identify either
UIR of $T$ as the trivial one.

\subsection { A trio of $\RP3$'s}  

\noindent
Now we revert to the special case where $\Gamma$ is abelian, meaning in
effect that it merely associates a sign with each ordered pair of primes.
We can represent the various such $\Gamma$ pictorially by drawing three
dots to represent the three primes and an arrow to represent each ordered
pair that receives a minus sign (meaning the a slide of the first prime
through the second produces a phase-factor of $-1$).  Each distinct diagram
of this type then gives rise to a different class of UIR's of $(slides)$,
and therefore furnishes a different building block for constructing UIR's
of $G$.

Perhaps the most interesting abelian UIR of $(slides)$ comes from the
cyclic graph in which the three dots and arrows form a circle.  Clearly
this pattern leaves $\Integers_3\subseteq(perms)$ as the unbroken subgroup
$K_0$, so we acquire three distinct UIR's of $G$, corresponding to the
three possible UIR's of $\Integers_3$.  What is remarkable here is first of
all the pattern of geon identity, which is expressed not by a permutation
group $\S_n$ at all, but by the cyclic group $\Integers_3$.  With this new
type of group comes a new type of statistics, in which a cyclic permutation
of the geons produces the complex phase $q$ or $\bar{q}$, $q=1^{1/3}$ being
a cube-root of unity.\footnote{*}%
{The same UIR of $\Integers_3$
 also occurs in connection with parastatistics, where
it represents only a proper subspace of the full state-space.  Indeed, any
UIR of any finite group can be realized in connection with parastatistics,
since any finite group is a subgroup of some permutation group.}


Although this pattern of identity is unusual for the simple type of
particle that physics usually deals with, it has obvious precedents in the
social world of human beings.  There it might happen, for example, that
three people could stand in a triangle so that each was the teacher (in
some different subject) of the one to his/her right.  The pattern would
then be preserved by a cyclic permutation, but not if two of the people
changed places while the third stayed put.

\subsection {More than three $\RP3$'s}  

\noindent
An interesting possibility in this case is that of ``projective
statistics'', meaning a type of statistics expressed by a properly
projective representation of the permutation group or one of its
subgroups.\footnote{$^\dagger$}%
{Unlike the ``cyclic statistics'' we just met, a projective statistics would
not be realizable in connection with parastatistics.}
We do not have an example yet, but there seems to be no good reason why one
shouldn't exist.  We would need at least four geons because $\S_n$
possesses properly projective representations only for $n\ge{}4$.  We would
also need a non-abelian UIR $\Gamma$ of $(slides)$, because in the contrary
case, the projective multiplier $\sigma$ will always be trivial.  Thus, the
simplest example one might try to construct, would employ a two dimensional
representation $\Gamma$, so chosen that the unbroken subgroup
$K_0\subseteq\S_4$ would come out as the subgroup $\S_4^{even}$ of even
permutations, and $\sigma$ would come out as the projective multiplier for
the spin=$1/2$ representation of the symmetry group of a regular
tetrahedron (to which $\S_4^{even}$ is isomorphic).

From the above examples it emerges clearly, we believe, that the
possibility of non-Euclidean spacetime topology introduces a once
unexpected richness into quantum gravity.  In particular it brings with it %
\BulletItem {topological particles (geons)}
\BulletItem {half integer spin in pure gravity}
\BulletItem {distinct quantum sectors (both continuous and discrete
            families of them)}
\BulletItem {quantum multiplicity}

\noindent
 We have not really discussed the second of these, and we did not
mention the third at all before now, but we list them here to help
illustrate the fertility of topology in the quantum context.  

We have seen, moreover, that the mathematical structure theorems for
representations of semidirect product groups provide a remarkably natural
physical description of the different sectors which can occur.  Indeed
these theorems read almost as if they had been expressly designed to
describe the representations in the language of quantum particles and their
properties!  In this language identical geon statistics is expressed by a
(possibly projective) unitary irreducible representation of the unbroken
subgroup of the group of permutations of identical primes.  Two noteworthy
features of the resulting interpretation are that
\BulletItem{All the familiar types of statistics can occur, together with
  some new, unexpected types.}
\BulletItem{New patterns of particle identity occur, in which not all
  permutations of the identical particle leave the physics invariant.}

\noindent
It is interesting that the most novel of these features are associated with
the slide diffeomorphisms, which correspond physically to processes in
which one geon ``slides thru'' another.  For this reason it seems possible
that analogous condensed matter effects could occur with objects like
vortices, which also can slide through one another in an obvious manner
[3].)

Beyond the existence of sectors associated with new types of particle
statistics and indistinguishability, it seems that continuous parameters
will also occur (we saw an example in the case of two $\RP3$ geons), so
that a great multiplicity of sectors can be expected to exist, even with a
given spatial topology.  This answers our initial question about new
topological parameters, and it seems that the answer is that there are many
--- probably too many in fact, since they can be chosen so that the
spin-statistics correlation fails, indicating\footnote{*}%
{If geons propagating in a flat ambient metric admit an approximate
description in terms of an effective flat-space quantum field theory, then
they must satisfy the standard spin-statistics correlation.}
that quantum gravity is more akin to a phenomenological theory than a
fundamental one.  One can expect that unfreezing the topology will remove
some of these unwanted sectors, but we would conjecture that a deeper,
discrete theory will be needed to restore a physically reasonable degree of
uniqueness to quantum gravity.

\bigskip\noindent
This research was partly supported by NSF grant PHY-9600620.

\ReferencesBegin

[1] 
  R.D.~Sorkin and S.~Surya,
``An Analysis of the Representations of the Mapping Class Group of a 
    Multi-geon Three-manifold'' 
   $\langle$e-print archive: gr-qc/9605050$\rangle$	

[2] 
 R.D.~Sorkin, 
``Introduction to Topological Geons'',
   in P.G. Bergmann and V. de Sabbata (eds.),
   {\it Topological Properties and Global Structure of Space-Time}, pp. 249-270
    (Plenum, 1986)

[3] A.P.~Balachandran, private communication.

\end